\documentclass[aps,pra,twocolumn,showpacs,preprintnumbers,amsmath,amssymb,longbibliography]{revtex4-2}
\usepackage{chngcntr}
\usepackage{amsmath,amssymb}
\usepackage{verbatim}
\usepackage{graphicx}
\usepackage{subfigure}   
\usepackage{float}

\usepackage[normalem]{ulem}

\usepackage[T1]{fontenc}
\usepackage[utf8]{inputenc}

\usepackage{xcolor}
\usepackage[colorlinks=true, linkcolor=blue, citecolor=blue, urlcolor=blue]{hyperref}

\usepackage[
    colorlinks=true,
    citecolor=blue,    
    linkcolor=blue,    
    urlcolor=blue      
]{hyperref}

\usepackage{etoolbox}



\makeatletter

\renewcommand\section{\@startsection{section}{1}{0pt}
  {0.8em plus 0.2em minus 0.1em}   
  {0.6em plus 0.1em minus 0.1em}   
  {\normalfont\bfseries\normalsize\centering}}   
\renewcommand\subsection{\@startsection{subsection}{2}{0pt}
  {0.6em plus 0.2em minus 0.1em}
  {0.4em plus 0.1em minus 0.1em}
  {\normalfont\normalsize\centering}}       
  
\makeatother
\begin{document}
\title{Klein bottle ratio of two-dimensional ferromagnetic Potts models}
\author{Zi-Han Wang}
\author{Li-Ping Yang}
\email{liping2012@cqu.edu.cn}
\affiliation{Department of Physics and Chongqing Key Laboratory for Strongly Coupled Physics, Chongqing University, Chongqing 401331, China}

\definecolor{burnt}{cmyk}{0.2,0.8,1,0}
\date{\today}

\begin{abstract}
The weakly first-order nature of the two-dimensional five-state ferromagnetic Potts model poses challenges for numerical study. Using density-matrix and tensor-network renormalization group methods, we investigate these transitions of the $q$-state Potts model via the Klein bottle ratio $g$ on original and dual lattices. Finite-size scaling of $g$ as a function of transverse system size $L_y$ accurately locates the critical points for $q = 3, 4, 5, 6$. We further examine the transfer-matrix spectra and entanglement entropy, extracting central charges through toroidal and Klein bottle boundary conditions. For $q = 5$, the extracted central charge ($c \approx 1.14811$) is close to the real part of the theoretical value $c_{5\text{-Potts}} = 1.1375 \pm 0.0211 i$ predicted by complex conformal field theories(CFTs). The observed drift in the scaling exponent $b$ effectively distinguishes the continuous transition from the weakly first-order regime. Furthermore, the extrapolated divergence of $g$ confirms the first-order character of the $q=5$ Potts model, while for the $q=4$ model, both the CFT-predicted eigenvalue distribution and the logarithmic corrections are numerically verified.

\end{abstract}

\maketitle

\section{Introduction}
The Potts model serves as a fundamental generalization of the Ising model, where each lattice site can take $q$ distinct spin states\cite{potts1952, onsager1944}. It is well-established that the two-dimensional $q$-state Potts model exhibits a continuous phase transition for $q \leq 4$, whereas a first-order phase transition emerges when $q > 5$\cite{Baxter1973Potts,wu1982}. In particular, the transition at $q=5$ is characterized as ``weakly first-order,'' where the correlation length $\xi$ becomes exceptionally large\cite{iino2019} compared to the lattice spacing and therefore exhibits an approximate scale invariance\cite{GRZ-SciPost, GRZ-jhep}. The renormalization group analysis reveals the essential singularity in latent heat, which renders the jumps of physical observables for $q\to 4^{+}$ almost undetectable in numerical simulations\cite{Nienhuis1979PRL,cardy1980, nauenberg1980,binder1981}.

At the critical point of continuous phase transition, the infinite correlation length leads to scale invariance, which in two dimensions is enhanced to conformal invariance. This allows for a rigorous description via conformal field theory (CFT){\cite{difrancesco1997}}. Recently, the complex CFTs describing a pair of complex fixed points uncovers the close relation to weakly first-order transitions in two-dimensional $q$-state Potts models\cite{GRZ-SciPost, GRZ-jhep}. The subsequent numerical simulation for $(1 + 1)$-dimensional quantum $q$-state Potts model demonstrates the ``shadow effect'' of these complex fixed points{\cite{ma2019}}. Furthermore, by incorporating non-Hermitian interactions (complex couplings), researchers have reclaimed this ``lost'' conformality by directly identifying the conjugate fixed points\cite{tang2024}. When $q=4$, two renormalization group (RG) fixed points collide and merge, which has already claimed in the early-stage study\cite{cardy1980,nauenberg1980}. It is intriguing that these fixed points move into the complex plane as a conjugate pair\cite{tang2024,jesper2024} for $q > 4$.
For the five-state Potts model, the complex central charge is estimated as $c_{5\text{-Potts}} \approx 1.138 \pm 0.021i$\cite{GRZ-SciPost}, characterizing its weak first-order nature within the framework of complex CFTs. This paradigm has also sparked renewed interest in Potts and related loop models\cite{Tang2025,Jacobsen2026,SciPostPhys.17.4.099,l1cp-6gzr}.

To probe systems where the computational size $L$ is significantly smaller than the correlation length $\xi$, finite-size-scaling (FSS) analysis is the standard tool for analysis{\cite{cardy1980, binder1984}}. For genuine continuous transitions, observables collapse onto a single scaling curve governed by universal critical exponents. By contrast, weakly first-order transitions---having a large but finite $\xi$---can exhibit pseudo-critical behavior and approximate conformal invariance\cite{prx-2017,ma2019,jesper2024} that mimics continuous scaling with effective exponents. This crossover regime therefore complicates the unambiguous classification of the transition order{\cite{iino2019, morita2019, fisher1982}}.

In numerical simulations of Potts models, magnetic susceptibility and the associated Binder cumulant are standard diagnostics for identifying phase transitions
\cite{iino2019, morita2019, Tseng2022}. For first-order transitions, the correlation-length exponent is characterized by the trivial value $\nu=1/d$\cite{prl-1975,fisher1982} (specifically $\nu = 1/2$ in two dimensions). Recent study\cite{morita2019}
demonstrates that the high-order tensor renormalization group (HOTRG)\cite{xie2012}
 method provides a more accurate fitting of $\nu$ for the five-state Potts model than conventional Monte Carlo results. By operating directly in the thermodynamic limit and leveraging the Klein bottle ratio $g$\cite{tu2017}, tensor network algorithms have successfully determined the critical properties of various systems, including q-state clock models\cite{li2020}.

In this article, we employ density-matrix\cite{white1992, white1993} and tensor-network renormalization group algorithms\cite{XT} to calculate the Klein bottle ratio $g$ {\cite{tu2017, zhang2023}} of two-dimensional ferromagnetic Potts models on square lattice. $g$ serves as a high-precision diagnostic for phase transitions in the Potts model, which is universal and only depends on the quantum dimensions of the primary fields\cite{tang2019}. Similar to the method and scheme in Ref.\cite{li2020}, $L_x\to \infty$ is implemented in the perspective of column-to-column transfer matrix, and thus $L_x \gg L_y$ are automatically guaranteed. $L_y$ takes values in range $8$-$70$ for comparison and fitting. We compute Klein bottle ratios for $q=3, 4, 5, \text{ and } 6$ to systematically follow the crossover from standard CFT behavior toward descriptions by complex CFTs.

The article is structured as follows: In Sec.\ref{MM}, We will introduce the Potts model, tensor network algorithms, and the Klein bottle ratio. In Sec.\ref{results}, the numerical results of different $q$-state Potts model are demonstrated and discussed in detail, especially in the cases of $q=4,5$. Finally, we go into the conclusions in Sec.\ref{conclusion}.

\section{Model and Simulation Methods}
\label{MM}

\subsection{Potts model}
The Hamiltonian of the $q$-state Potts model can be written as
\begin{equation}
H = -J \sum_{\langle i j \rangle} \delta_{s_{i}, s_{j}} .
\label{eq:PottsHamiltonian}
\end{equation}

Here, $s_i$ denotes the spin variable on site $i$, taking values $1,2,\ldots,q$; the Hamiltonian sums over the nearest-neighbor pair denoted by $\langle ij\rangle$ ;  $J$ is set to $1$ to discuss the ferromagnetic interaction. $\delta$ is the Kronecker delta function, which equals 1 if $s_i = s_j$ and 0 otherwise;  

The two-dimensional Potts model holds $S_q$ permutation symmetry due to the fact that $\delta_{s_{i}, s_{j}}=\delta_{P(s_{i}), P(s_{j})}$ under any permutation $P\in S_q$, which leads to the classification of Potts critical lines\cite{Cardy1996, delfino2017}. In case of $q=4$, there is the marginal term in the conformal field theory description at the critical point\cite{nauenberg1980,cardy1980}. As a result, the critical scaling is no longer purely power-law but acquires characteristic multiplicative logarithmic corrections{\cite{cardy1980,nauenberg1980,salas1997}}.

In the five-state Potts model the correlation length is extremely large---estimated to be about 2,500 lattice spacings\cite{iino2019}---far exceeding typical computational system sizes. Consequently, we employ finite-size scaling to analyze the system\rq s phase transition. In a true first-order transition the correlation length remains finite and conformal field theory does not apply; therefore physical observables need not obey simple finite-size scaling forms and generally do not collapse onto a single curve. However, for the five-state Potts model the anomalously large correlation length can produce pseudo-critical scaling in data-collapse analyses, yielding apparent nontrivial critical exponents\cite{GRZ-SciPost, GRZ-jhep, iino2019}.
\subsection{Tensor construction}

\begin{figure}[t]
    \centering
    \includegraphics[width=\linewidth]{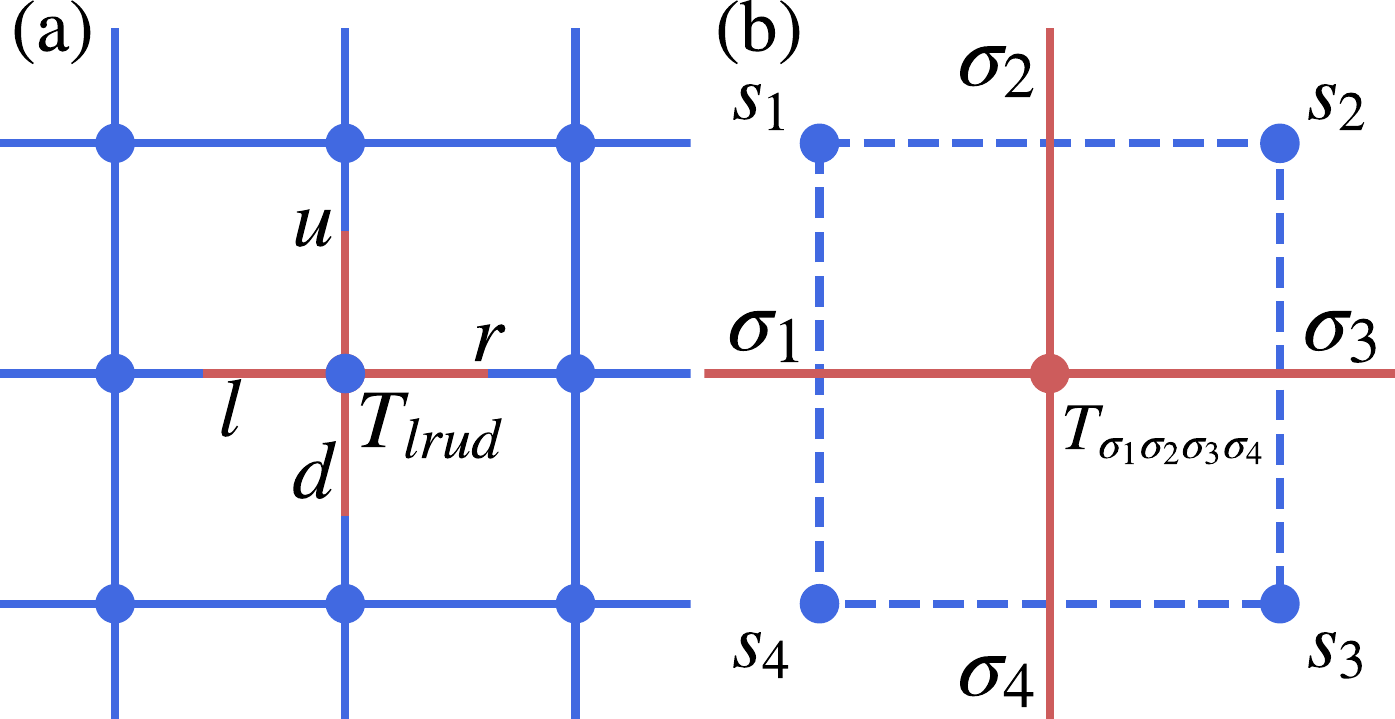}
\caption[
]{
Tensor-network representation of the partition function for the $q$-state Potts model on a square lattice:(a) Original lattice; (b) Dual lattice. The dual spin variable $\sigma_{\alpha}$ is defined on each bond and takes a value depending on whether the two Potts spins at the ends of the bond are equal. For example, $\sigma_{i} = \text{mod}(s_i-s_{i-1}+q,q)$, where $s_i$ denotes a Potts spin and $s_0=s_4$.
}
\label{fig:tensor-construction}
\end{figure}

Within the tensor-network framework, the partition function of the $q$-state Potts model can be expressed as a contraction of local tensors $T${\cite{li2020,wang2014}}:
\begin{equation}
Z = \sum_{\cdots l r u d\cdots} (\cdots T_{l r u d} \cdots) .
\label{eq:Ztensor}
\end{equation}

For the original lattice, the local tensor at each vertex (denoted by blue solid point in Fig.~\ref{fig:tensor-construction}) of the square lattice, is defined as
\begin{equation}
T_{l_{i}, r_{i}, d_{i}, u_{i}}
  = \frac{\sqrt{ I_{l_{i}} I_{r_{i}} I_{d_{i}} I_{u_{i}} }}{q}\,
    \delta_{\mathrm{mod}(l_{i}-r_{i}+d_{i}-u_{i},\, q)} .
\label{eq:TiTensor}
\end{equation}
Here, $l_i,r_i,d_i,u_i$ denote the bond indices emitting from site $i$ to (left, right, down, up) directions.
And $I_m = \sum_{\theta} e^{-im\theta}\, e^{\beta\delta_{\theta,0}}$ with $\theta = 2\pi n/q$,  
and each tensor index ($l,r,u,d,m$) and $n$ run from 1 to $q$.

In the case of the dual lattice, the local tensor is defined at the centers of the original lattice, forming another square lattice. The local tensor takes the form
\begin{equation}
\small
T^{\mathrm{dual}}_{\sigma_1\sigma_2\sigma_3\sigma_4}
 = q\!\sum_m\!
   \sqrt{J_{\sigma_1}J_{\sigma_2}J_{\sigma_3}J_{\sigma_4}}\!
   \delta_{\mathrm{mod}(\sigma_1{+}\sigma_2{+}\sigma_3{+}\sigma_4,q)} .
\end{equation}

Here,
$J_m = e^{\beta \delta_{m,0}}$,  
so that $J_{\sigma}=e^{\beta}$ for $m=0$ and $J_m=1$ otherwise. The tensor unit construction of original and dual lattices are illustrated in Fig.~\ref{fig:tensor-construction}.

When there exhibits only one critical point, the singularity of the partition function apply to the tensor-unit construction of the original lattice and dual lattice\cite{Mussardo,XT}. The inverse transition temperature $T_c = 1/\beta_c$ ($k_B=1$) is given by\cite{Chenjing-cpl,XT}
\begin{equation}
\frac{I_m(\beta_c)}{I_0(\beta_c)}=\frac{J_m(\beta_c)}{J_0(\beta_c)}.
\end{equation}
Therefore, $T_c = 1/ \ln(1+\sqrt{q})$, same as the theoretical result\cite{wu1982}.

\subsection{Klein Bottle Ratio $g$}

In critical quantum systems defined on a Klein bottle, the ratio $g$ directly arises from the manifold's non-orientability, representing a universal ``ground-state degeneracy''\cite{Affleck1991}. The Klein twist procedure is devised to extract the Klein bottle ratio $g$ by the ratio of the partition functions on the Klein bottle and torus\cite{tu2017}. Consequently, the numerical simulation are developed for the calculation of $g$\cite{tang2017,wang2018b,li2020,ueda2026}. This universal parameter $g$ applies to not only in critical quantum systems but also in critical statistical classical systems.
\begin{equation}
g = \frac{Z^{K}(2L_x,\, L_y/2)}{Z^{T}(L_x,\, L_y)} .
\label{eq:KleinBottleRatio}
\end{equation}
where $Z^{K}$ and $Z^{T}$ are the partition functions on the Klein bottle and torus.  
$L_x$ and $L_y$ are the length of the system in $x$ and $y$ directions, with the requirement $L_x\gg L_y$.

\begin{figure}[t]
    \centering
\includegraphics[width=0.79\linewidth]{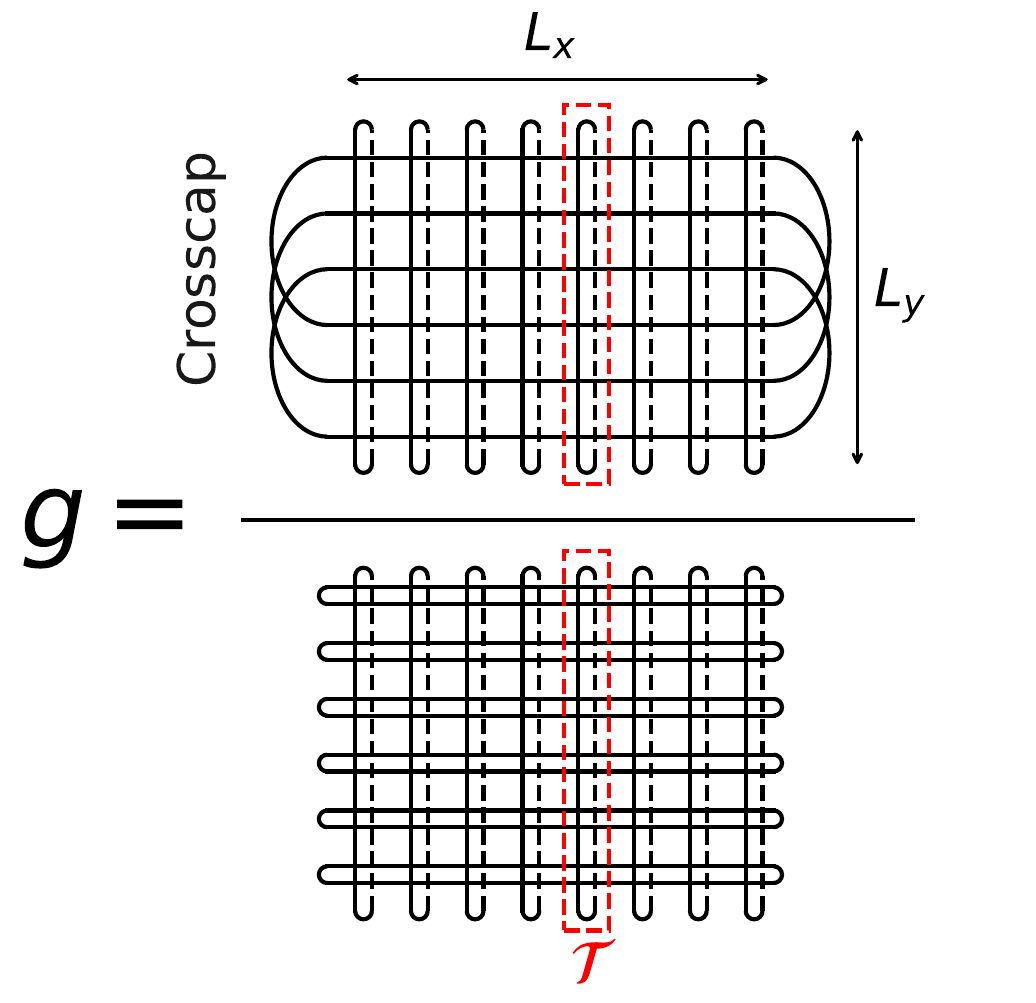}
\vspace{0em} 
    \caption{
    The calculation of the Klein bottle ratio $g$ is implemented by the ratio of two tensor network contractions. Along $L_y$ direction, the crosscap contraction is adopted. $\mathcal{T}$ means the column transfer matrix.
    }
    \label{fig:ZK/ZT}
\end{figure}

For the four-state Potts model, the universal class corresponds to a $Z_2$ orbifold of a compactified boson CFT with the radius $R=2\sqrt{2}$ and the central charge $c=1$. 

The partition function of the torus and the Klein bottle can be expressed in the following form
\begin{equation}
\begin{aligned}
\ln Z^{T} &\simeq -f_{0} L_{x} L_{y} + \frac{\pi c}{6 L_{y}} L_{x}, \\[4pt]
\ln Z^{K} &\simeq -f_{0} L_{x} L_{y} + \frac{\pi c}{24 L_{y}} L_{x} + S_{\mathrm{KB}},
\end{aligned}
\label{eq:ZT-ZK}
\end{equation}
where $f_0$ is a non-universal constant and $c$ is the central charge. $S_{\mathrm{KB}}=\ln g$ is known as the Klein bottle entropy, which is universal. By cutting-flipping-sewing operations\cite{tang2017}, Fig.~\ref{fig:ZK/ZT} realizes the calculation of the Klein bottle ratio $g$: the ratio of the two partition functions with crosscap and periodic boundaries along $L_y$, respectively\cite{tu2017,tang2017}. Furthermore, $g$ can be mapped to a boundary entropy, which obeys universal scaling relations at continuous phase transitions. 
To ensure the same column transfer matrix denoted as $\mathcal{T}$ in  the numerator and denominator, the left-right reflection symmetry of the local tensor is demanded. The details can be found in Ref.\cite{sm-li2020}. The further study shows that, not only at the critical point but also in its vicinity, the Klein bottle entropy obeys universal scaling laws, allowing it to characterize the critical behavior of the system.{\cite{chen2017b,lu2001,tan2025,zhang2023,tang2020}}

The partition function of Klein bottle and torus can be strategically designed within the tensor-network framework{\cite{wang2018b,li2020,li2021}}. As is shown in Fig.~\ref{fig:ZK/ZT}, $Z^T=\text{Tr}(\mathcal{T}^{L_x})$, the DMRG algorithm can be applied in solving the dominant eigenvector of $\mathcal{T}$ with the periodic boundary.
In the numerical simulations, $L_x$ is taken in the thermodynamic limit, which is encoded in the algorithms, and $L_y$ varies up to $70$.

The dominant eigenvalue $E_{max}$ and the corresponding eigenvector $|\psi_{max}\rangle$ of $\mathcal{T}$ can be obtained by using the DMRG algorithm\cite{white1992,white1993}. As a result, $Z^{T}=\text{Tr}(\mathcal{T}^{L_x})\approx E_{{max}}^{L_x}$. The expression is exact when there is not degeneracy for $E_{max}$ when $L_x \to \infty$. In practice, we can impose a crosscap contraction on $|\psi_{max}\rangle$, i.e., $g=|\langle\psi_{max}|\mathcal{C}_{\mathrm{latt}}\rangle|^2$\cite{zhang2023,2601.21502}. Consequently the Klein bottle ratio $g$ and $S_\text{KB}$ can be obtained. {\cite{li2020,wang2018b,tang2020,2601.21502}}.

\section{Numerical Results}
\label{results}

\begin{figure}[t]
    \centering
    \includegraphics[width=0.95\linewidth]{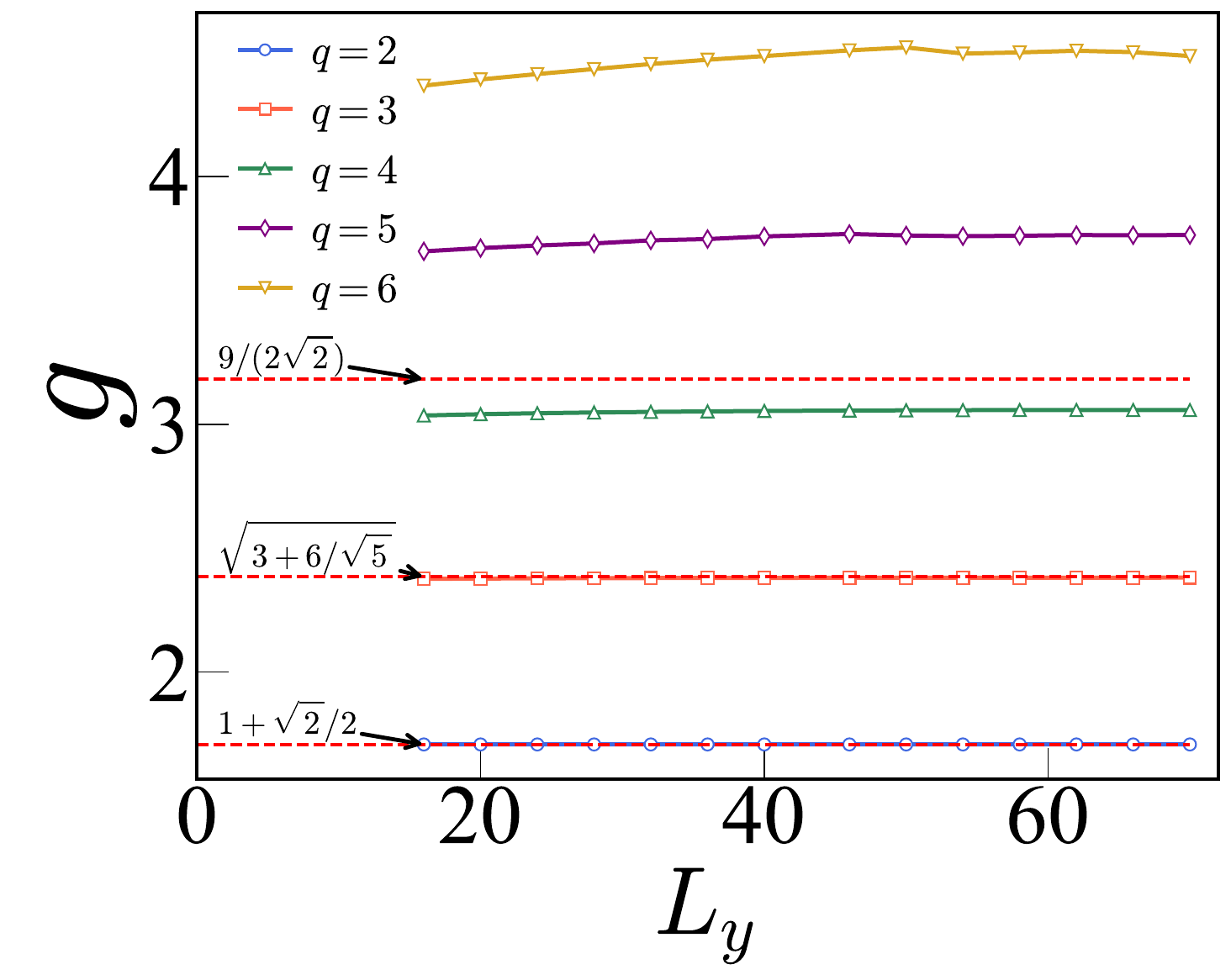}
    \caption{
    The Klein bottle ratio $g$ at critical temperature $T_c$ for the $q$-state Potts models ($q=2,3,4,5,6$). $L_y$ ranges from $16$ to $70$ in steps of $4$.  The truncation dimension $D=200$. The dash lines from CFT description are drawn for reference.
    }
    \label{fig:KBR_all}
\end{figure}

The Klein bottle ratio $g$ at the critical temperature is computed for systems with different $L_y$. As is shown in Fig.~\ref{fig:KBR_all}, for $q=2$,  $g$ is a constant which reaches the theoretical limit even for very small $L_y$ . For $q=3$, the Klein bottle ratio remains essentially constant over the entire range of system sizes and quickly saturates to its theoretical value. For $q=4$, the boundary entropy saturates to a value slightly below the CFT prediction. This discrepancy is likely caused by a marginally irrelevant term{\cite{tang2017}}. The Klein bottle ratio shows a weak but noticeable upward shift as $L_y$ increases. The up-shifting trend becomes more prominent in the cases of $q=5,6$, and a refined and detailed calculation demonstrates the finite-size effect, as shown in Fig.~\ref{D56}.

At the critical point of the four-state Potts model, the theoretical Klein bottle entropy is  
$S_{\mathrm{KB}} = \ln\!\left(\frac{9}{2\sqrt{2}}\right) \approx 1.1575$\cite{tang2017}. Further calculation shows that the Klein bottle ratio $g$ converges well with respect to $L_y$ by increasing $D$ from $200$ to $350$.
For the four-state Potts model, $g$ exhibits a very weak finite-size drift with increasing $L_y$. As shown in Fig.~\ref{fig:g-logLy}(b), incorporating $\ln L$ logarithmic corrections induced by the marginally irrelevant operator markedly improves the extrapolation, yielding $g = 3.203$ ($S = \ln g \approx 1.1641$). This result achieves excellent agreement with the exact CFT prediction $g = 3.182$ ($S \approx 1.1575$), demonstrating that $g$ serves as a robust diagnostic even in the presence of marginal logarithmic corrections.

For the five-state model, the Klein bottle ratio reaches $g\!\approx\!3.8049$ ($S=\ln g\!\approx\!1.336$). When $q>4$, the Klein bottle ratio $g$ exhibits a significant size-dependent effect. 
\begin{figure}[t]
    \centering
    \includegraphics[width=\linewidth]{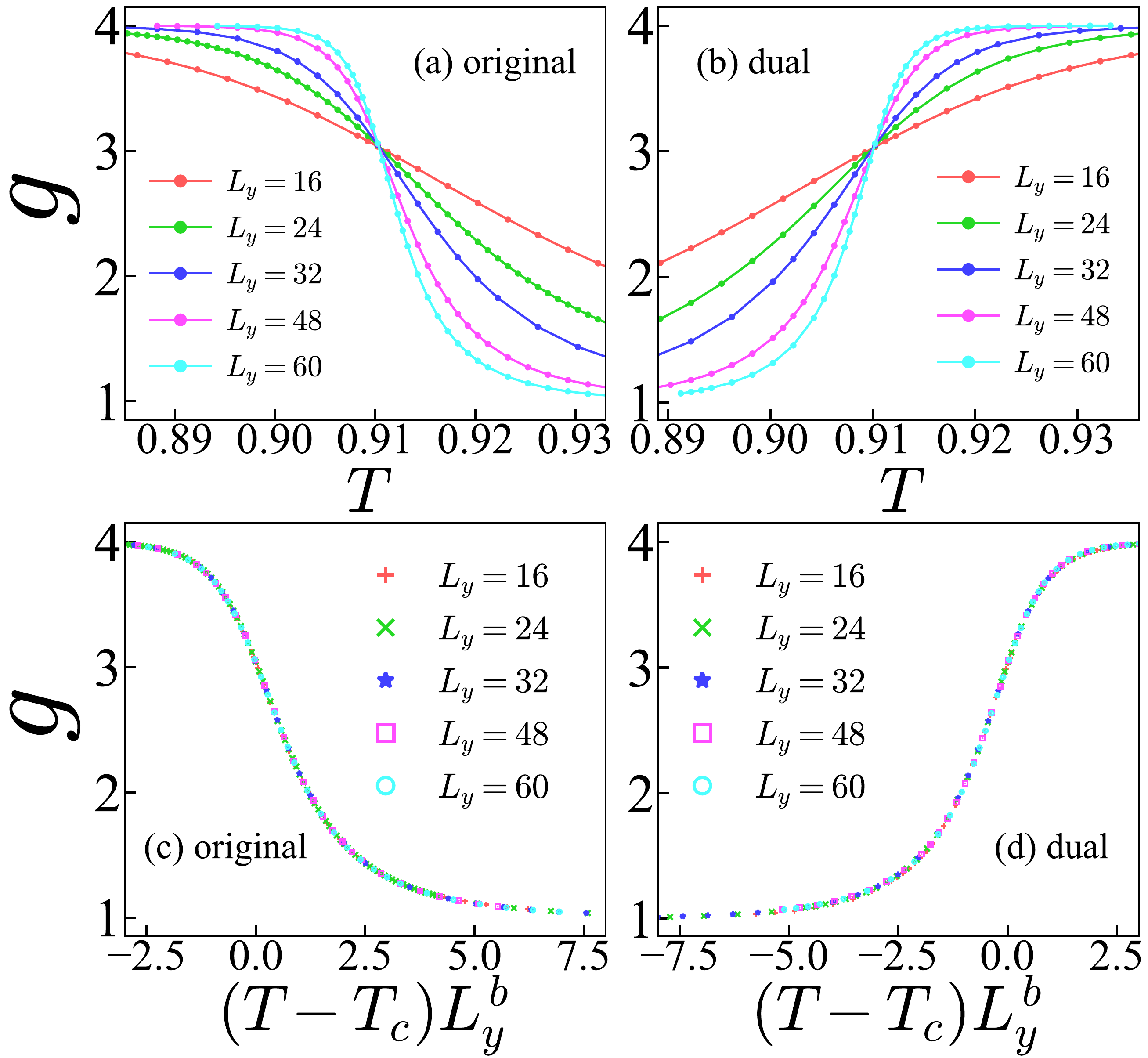}
\caption{
\textbf{$q=4$ Potts model.}Temperature dependence of the Klein bottle ratio $g$ with the length $L_y$ ranging from $16$ to $60$. 
(a) original lattice; (b) dual lattice; (c) data collapse on the original lattice: $T_c = 0.91035$, $b = 1.39581$; 
(d) data collapse on the dual lattice: $T_c = 0.91011$, $b = 1.36798$.}
    \label{fig:q4merged}
\end{figure}
\subsection{ Critical behaviors of $q=4,5$}

Fig.~\ref{fig:q4merged} illustrates the temperature dependence of the Klein bottle ratio $g$ for $q=4$ on both original (a) and dual (b) lattices. The curves for different $L_y$ have a common intersection point, which corresponds to the phase-transition critical point. Clearly, the transition becomes sharper with  $L_y$ increasing in the vicinity of the critical region. For the original lattice, $g$ monotonically decreases from 4 (ordered phase) to 1 (disordered phase) as the temperature increases.  While the behavior reverses the change on the dual lattice. As panels (c) and (d) demonstrate, data collapse is achieved by rescaling the temperature axis with $(T-T_c)L_y^b$. Utilizing Gaussian-process regression{\cite{harada2015,harada2011}}, the optimal fitting parameters for both lattices are obtained. 

For the four-state Potts model, the primary field driving the phase transition is the thermal field $\varepsilon$ with conformal weight $h = \bar{h} = 1/4$ \cite{Dijk1989,Caselle1999,Chatelain2026}, which implies 
$b = 2 -2h = 3/2$ \cite{zhang2023}.
Our numerical simulations yield $b \approx 1.4$, deviating from the theoretical value $1.5$. 

Here, finite-size scaling requires careful treatment due to the presence of marginally irrelevant operators. Initially, a direct power-law extrapolation to the thermodynamic limit $L_y \to \infty$ yields $g \approx 3.1232$ ($S = \ln g \approx 1.1389$), as illustrated in Fig.~\ref{fig:g-Ly}(b). While this value is already closer to the exact CFT prediction ($g = 3.182$, $S \approx 1.1575$) than the previously reported Monte-Carlo result ($S \approx 1.108$ \cite{tang2017}), a noticeable discrepancy remains. To properly account for the logarithmic corrections and the non-universal initial condition of the RG flow \cite{tang2017}, we adopt a logarithmic ansatz $g(L_y) \sim a/(\ln(L_y)+b)$. As demonstrated in Fig.~\ref{fig:g-logLy}(b), this logarithmic fitting yields a marked improvement, shifting the extrapolated value to $g = 3.203$. This remarkable agreement with the exact CFT benchmark directly confirms the crucial role of marginally irrelevant operators in the four-state Potts model, while highlighting the high precision of $g$ as a critical probe.

Furthermore, we conduct independent data fitting for Klein bottle ratios with smaller and larger system sizes, respectively. The results show that good data collapse results can be achieved in both cases, providing the critical temperature $T_c=0.91035$ on original lattice and $T_c=0.91011$ on dual lattice, very close to the exact value $T_c=1/\ln(3) \approx 0.91024$.


\begin{figure}[t]
    \centering
    \includegraphics[width=\linewidth]{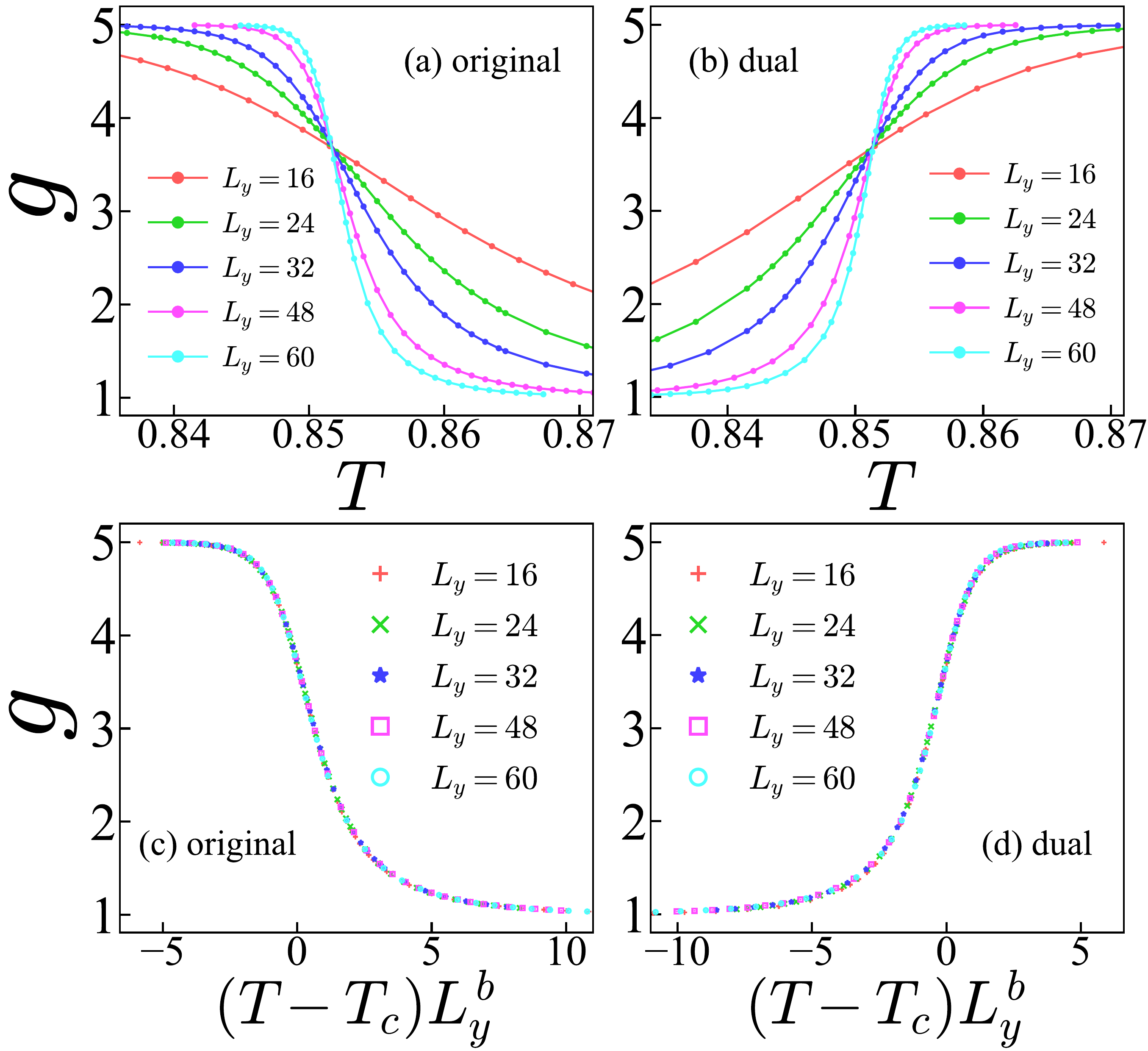}
    \caption{
    \textbf{$q=5$ Potts model.} Temperature dependence of the Klein bottle ratio $g$  with the length $L_y$ ranging 
from $16$ to $60$. 
(a) original lattice; 
(b) dual lattice; 
(c) data collapse on the original lattice, giving $T_c = 0.85167$, $b = 1.59605$; 
(d) data collapse on the dual lattice, giving $T_c = 0.85140$, $b = 1.57144$.
    }
    \label{fig:q5merged}
\end{figure}

As shown in Fig.~\ref{fig:q5merged}, the result of $q=5$ closely resembles that of $q=4$. The theoretical critical temperature is $T_c=1/{\ln(1+\sqrt{5})}${\cite{baxter2007}} $\approx 0.85153$. The data collapse leads to the consistent critical temperature $T_c=0.85167$ on original lattice and $T_c=0.85140$ on dual lattice, respectively. By virtue of Kramers–Wannier duality mapping high- and low-temperature regimes\cite{difrancesco1997}, numerical critical temperatures from the original and dual lattices naturally straddle the exact theoretical value.

In the cases of $q=4,5$, the difference of $T_c$ and $b$ from the data collapse between the original and dual lattice are roughly the same numerically. Further analysis shows that the fitting exponent $b$ is much more sensitive to the system size $L_y$, exhibiting the prominent scaling exponent drift for $q=5$, which is absent in $q=4$.

\subsection{Scaling exponent drift}
\begin{figure}[th]
\centering

\begin{minipage}{\linewidth}
\centering
\includegraphics[width=\linewidth]{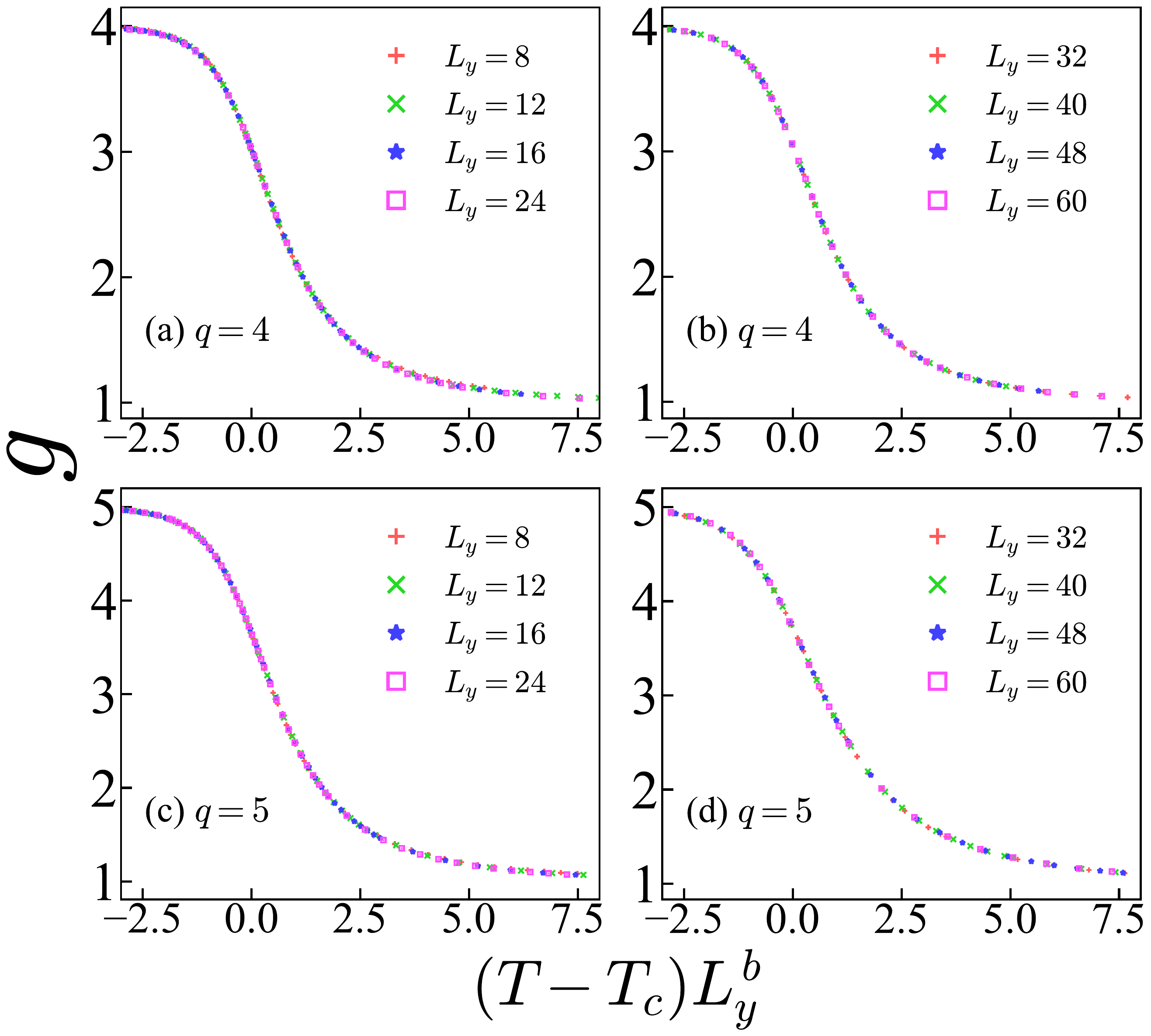}
\label{se}
\end{minipage}
\hfill
\caption{
 Data collapse of the Klein bottle ratio $g$ for the four- and five-state Potts model on original lattice with truncation dimension $D=280$. Smaller and larger sizes and the corresponding collapse are compared. $q=4$ case:
(a) the fitting of smaller-size systems shows $T_c = 0.91060$, $b = 1.39857$; 
(b) the fitting of larger-size systems shows $T_c = 0.91031$, $b = 1.40088$.
$q=5$ case: (c)  the fitting of smaller-size systems of five-state Potts model shows $T_c = 0.85198$, $b = 1.55653$; 
(d) the fitting of larger-size systems of five-state Potts model shows $T_c = 0.85163$, $b = 1.61904$.
}
\label{b-drift}
\end{figure}

Two $L_y$ subsets are chosen for comparison: the smaller one with $L_y = 8,12,16,24$, and the larger one with $L_y = 32,40,48,60$. In the four-state Potts model, the fitted $b = 1.39857$ for the smaller $L_y$ subset and $b=1.40088$ for the larger $L_y$ subset, respectively. In contrast, the data for the $q=5$ Potts model exhibit a systematic drift: the effective exponent increases from $b = 1.55653$ for smaller system sizes to $b = 1.61904$ for larger system sizes. 
Moreover, applying the exponent $b$ fitted from the smaller $L_y$ subset to the larger $L_y$ subset leads to a pronounced breakdown of data collapse. The curves visibly separate, whereas the data for $q=4$ collapse closely onto a single master curve. This contrast indicates that the apparent scaling observed in the $q=5$ model is merely pseudo-critical, persisting only up to a finite length scale. Beyond this scale, the data deviate from a universal scaling function, fully consistent with the weakly first-order nature of the five-state Potts model.

Fig.~\ref{b-drift} illustrates the results, with the left panels representing smaller system sizes and the right panels for larger ones. Despite the data exhibiting good collapse across all cases, a nuanced picture emerges concerning the critical parameters. Fitting the larger $L_y$ subset yields more accurate $T_c$ values for both the four- and five-state Potts models. The discrepancy in the critical exponent, $\Delta b = 0.06251$ for the five-state Potts model, is more than an order of magnitude larger than that for the four-state Potts model ($\Delta b = 0.00231$). The exponent $b$ exhibits a pronounced drift, which is consistent with the drift of the central charge and scaling dimensions reported for the one-dimensional quantum five-state Potts model in Ref.\cite{ma2019}. Specifically, our results for the five-state model reveal a systematic and robust increase in $b$ as $L_y$ increases. Taken together, these findings strongly indicate that the five-state Potts model undergoes a weakly first-order phase transition.

Although an intermediate value of $b \approx 1.59605$[Fig.~\ref{fig:q5merged} (c)] or $b \approx 1.57144$ [Fig.~\ref{fig:q5merged} (d)] can still facilitate a reasonable collapse across the entire data range, the prominent behavior of the exponent drift is absent in the $q=4$ case, which is closely related to the ``walking '' behavior in the five-state Potts model and thus drifting scaling dimension\cite{GRZ-SciPost}.

\begin{figure}[t]
    \centering
    \begin{minipage}{0.99\linewidth}
\centering
\includegraphics[width=0.99\linewidth]{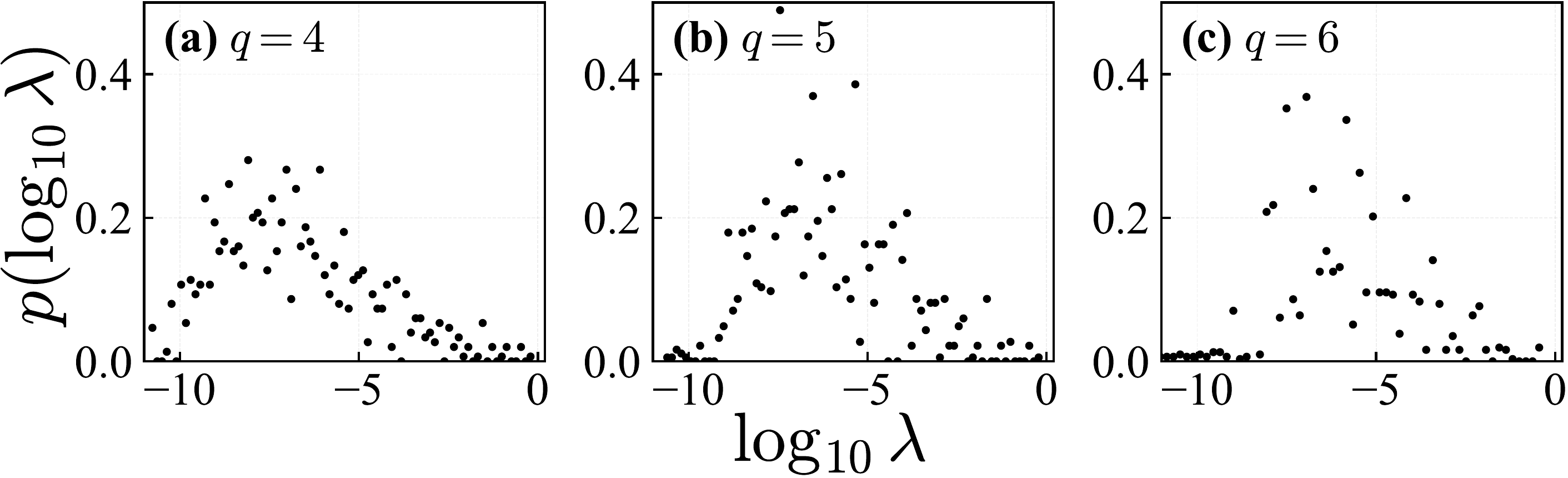}
    \caption{
    Spectrum distribution of the retained non-zero singular values $\lambda_i$ for the (a) $q=4$, (b) $q=5$ and (c) $q=6$ Potts models on the original lattice, obtained with a maximum bond dimension $D=280$ and $L_y=20$. 
    }
    \label{fig:spectrum}
    \end{minipage}
\end{figure}

\begin{figure}[!t]
    \centering
    \includegraphics[width=0.95\linewidth]{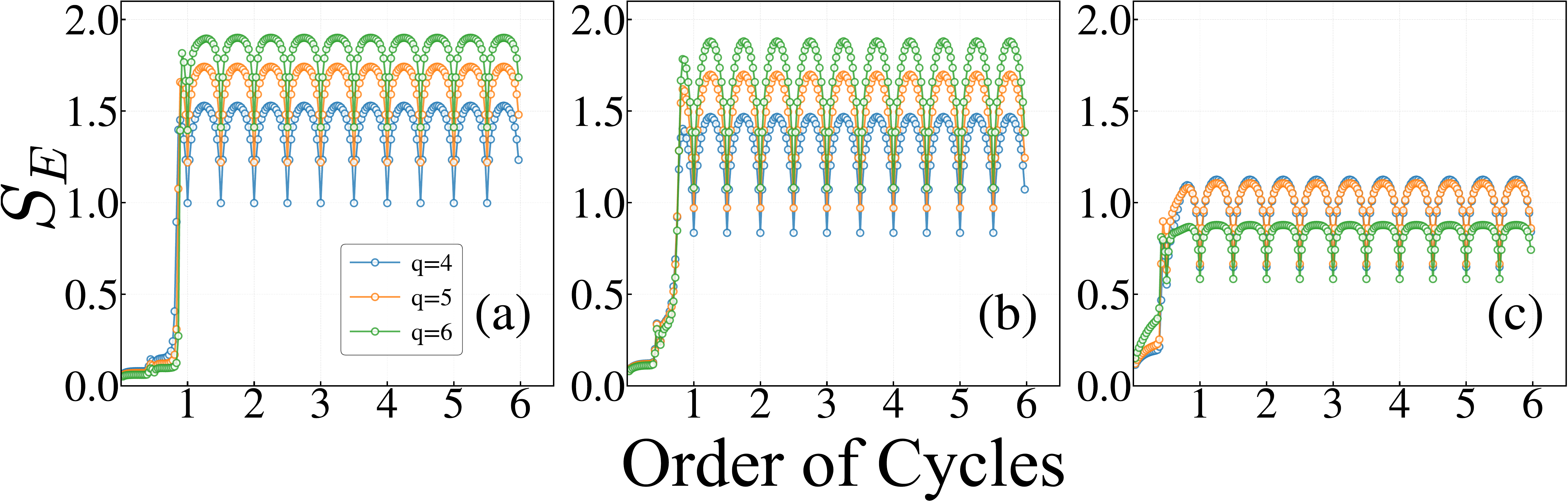}
    \caption{
     Evolution of the von Neumann entanglement entropy $S_E$ as a function of the order of cycles for with  $q = 4, 5,$ $6$ during the DMRG sweeping. The results are obtained for system size $L_y = 20$ and bond dimension $D = 280$. The three panels (a) (b) (c) correspond to different temperatures relative to the critical $T_c$ with (a) $T = T_c - 0.02$, (b) $T = T_c$, and (c) $T = T_c + 0.02$. 
    }
    \label{fig:EE}
\end{figure}

\subsection{ Spectrum of the transfer matrices }
Density-matrix and tensor-network renormalization group algorithms, in essence, are entanglement-based algorithms.  The entanglement is encoded in the singular-value-decomposition (SVD)  spectrum. The partition function of the torus is given by $Z^T = \mathrm{Tr} (\mathcal{T}^{L_x})$. The SVD spectrum $\{\lambda_i\}$ is obtained during the DMRG algorithm for the transfer matrix $\mathcal{T}$. $\{\lambda_i\}$ is arranged in descending order for truncation. $D$ is the retained bond dimension, i.e., the number of $\lambda_i$. In the finite-chain algorithm\cite{white1993}, there is $Dq$ singular values at each step of iteration. 

Fig.~\ref{fig:spectrum} demonstrates the spectrum distributions at the critical temperature in the cases $q=4,5,6$ before truncation, respectively. The two middle points along $y$- direction are chosen for the observation of the spectrum. The detailed analysis shows that the leading eigenvalues decay rapidly despite exhibiting high near-degeneracies, while the subsequent long-tailed distribution remains characterized by substantial quasi-degeneracy, which necessitates a larger bond dimension to maintain numerical precision.

The SVD spectrum can be used to calculate the von Neumann entanglement entropy $S_E=-\sum_i\lambda_i^2 \ln (\lambda_i^2)$. As shown in Fig.~\ref{fig:EE}, $S_E$ exhibits periodic oscillations with the DMRG sweeping over the periodic finite-chain. $S_E$ peaks at the middle of the chain during the variational optimization, and drops to a minimum at the boundary. It is remarkable that the entanglement entropy $S_E$ increases with the optional states of $q$ in the critical point and the left vicinity, however, when the temperature exceeds $T_c$, $S_E$ of $q=6$ is significantly smaller than $q=4,5$ cases. In the vicinity of the critical point, a striking similarity is observed between $q=4$ and $5$ cases, which is originated from the conformal criticality and approximate conformal criticality\cite{GRZ-SciPost,GRZ-jhep}. In sharp contrast,  as is shown in Fig.~\ref{fig:spectrum}, $q=5$ is much similar to $q=6$. It implies the first-order transition nature of five-state Potts model. The combination of Fig.~\ref{fig:spectrum}
and Fig.~\ref{fig:EE} demonstrate the ``weakly'' first order nature of the two-dimensional ferromagnetic five-state Potts model on square lattice.

\begin{figure}[t]
   \centering
    \begin{minipage}{0.98\linewidth}
        \includegraphics[width=\linewidth]{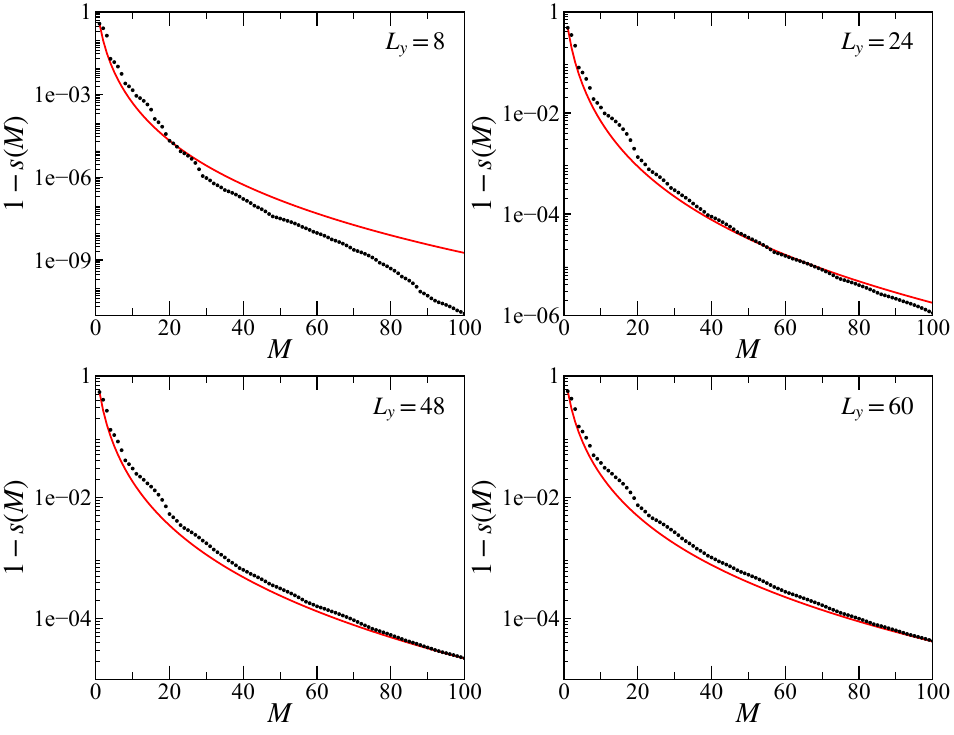}
    \end{minipage}
\caption{Cumulative sum of the lowest $M$ eigenvalues, $1-s(M)$, plotted against $M$ (up to $M=100$) for the four-state Potts model across various system widths $L_y = 8, 24, 48,$ and $60$ (black dots). The solid red line denotes the CFT prediction.}
\label{fig:sM}
\end{figure}

\begin{figure}[t]
   \centering
    \begin{minipage}{0.98\linewidth}
        \includegraphics[width=\linewidth]{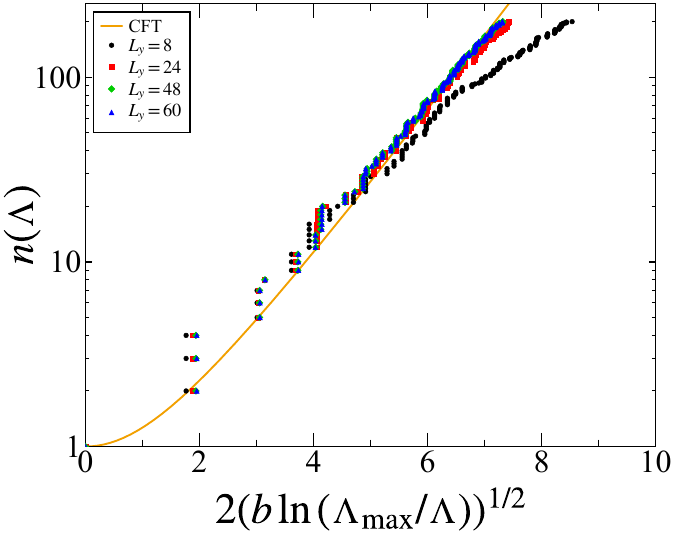}
    \end{minipage}
\caption{The cumulative count of eigenvalues $n(\Lambda)$ exceeding a given threshold $\Lambda$, plotted against the scaling variable $2\sqrt{b\ln(\Lambda_{\max}/\Lambda)}$ for system sizes $L_y = 8, 24, 48,$ and $60$ in the four-state Potts model. Here, $b=-\ln\Lambda_{\max}$. The solid curve denotes the CFT prediction.}
\label{fig:nLam}
\end{figure}

To verify the applicability of the conformal prediction \cite{Calabrese2008} to the four-state Potts model, we analyzed the cumulative eigenvalue spectrum $s(M) = \sum_{i=1}^{M} \Lambda_i$ (Fig.~\ref{fig:sM}). Here, $\Lambda_i = \lambda_i^2$ connects the eigenvalues of the reduced density matrix to the singular values $\lambda_i$ obtained from the DMRG algorithm, with $s(M)$ characterizing the majorization of the spectrum. As shown in Fig.~\ref{fig:sM}, the numerical results exhibit excellent agreement with the CFT prediction, which systematically improves as the system width $L_y$ increases. The distribution of the entanglement spectrum is further manifested in the functional dependence of $n(\Lambda)$ on $\Lambda$, where $n(\Lambda)$ represents the mean number of reduced density matrix eigenvalues exceeding a threshold $\Lambda$. As shown in Fig.~\ref{fig:nLam}, the numerical data across various system widths $L_y$ collapse remarkably well onto the theoretical curve predicted by CFT\cite{Calabrese2008}. 

As a consequence, achieving accurate and well-converged numerical results requires retaining a sufficiently large bond dimension $D$. Moreover, increasing the Potts parameter $q$ or the system size $L_y$ leads to a rapid growth in the required bond dimension for sufficient precision. Since the computational cost scales as $\mathcal{O}(D^3q^3)$, maintaining a sufficiently large truncation dimension $D$ becomes crucial for reliable calculations, especially for large $q$ and extended systems with large $L_y$. From the profile, $q=5$ takes more resemblance with $q=6$ other than $q=4$.

\begin{figure}[t]
   \centering
    \begin{minipage}{0.98\linewidth}
        \includegraphics[width=\linewidth]{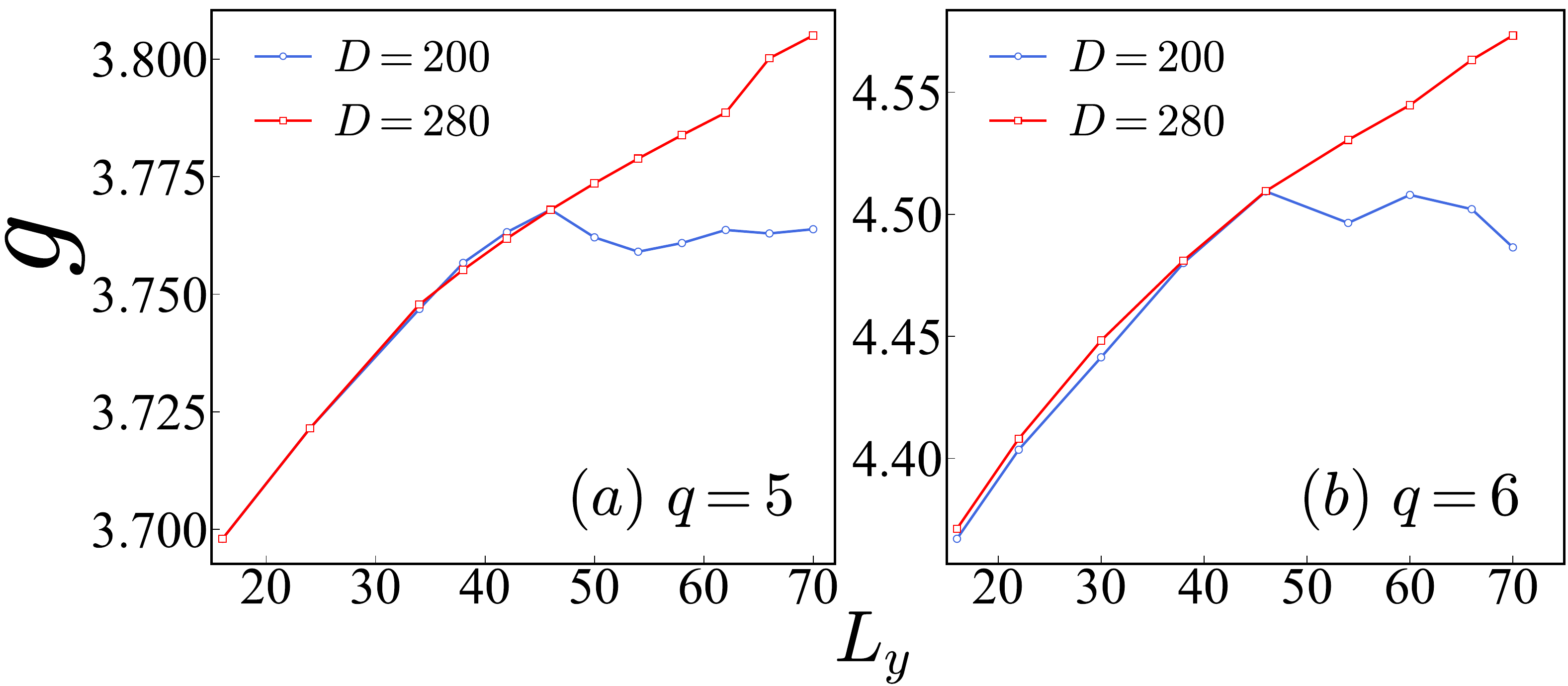}
    \end{minipage}
\caption{ Effects of bond dimension of the Klein bottle ratio $g$ on the original lattice at the critical temperature $T_c$ for the $q=5$ and $q=6$ Potts models.
 $D=200,280$ are compared on the curves of $g$ vs $L_y$.
}
\label{D56}
\end{figure}

\begin{figure}[t]
   \centering
    \begin{minipage}{0.98\linewidth}
\includegraphics[width=\linewidth]{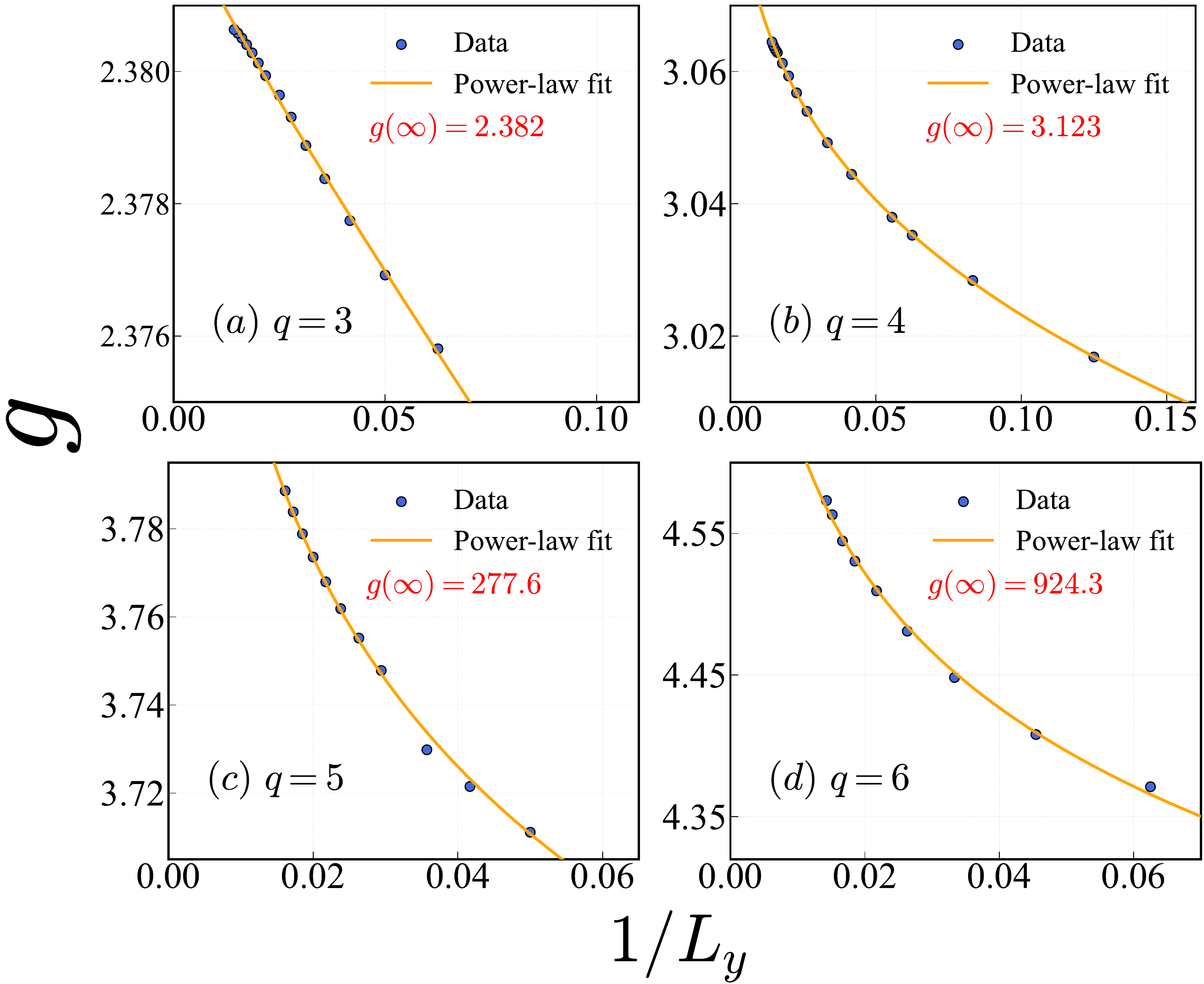}
\end{minipage}
\hfill

\caption{The Klein bottle ratio $g$ of the $q$-state Potts models ($q=3, 4, 5, 6$) versus {\boldmath$1/L_y$}. The blue dots represent the numerical results with a truncation dimension $D=280$. A power-law fitting is performed on $1/L_y$ as follows: $g = g(\infty) + a \times (1/L_y)^{-b}$.}
\label{fig:g-Ly}
\end{figure}

\begin{figure}[t]
\centering
\includegraphics[width=0.98\linewidth]{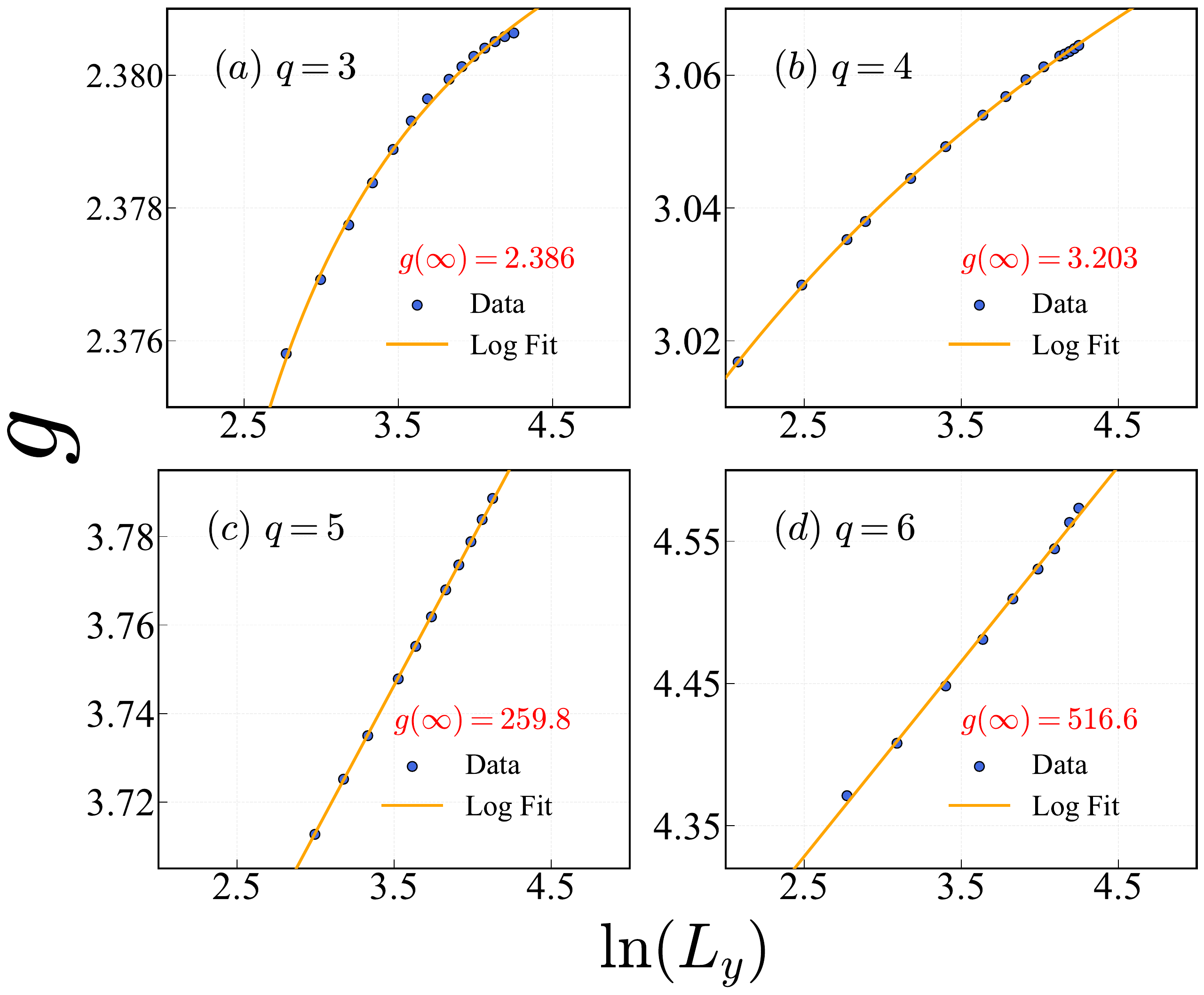}
\caption{Compared to Fig.~\ref{fig:g-Ly}, the fitting 
horizontal axis is {\boldmath$\ln(L_y)$}, the same data set applies. A fitting with the logarithmic correction is performed on $\text{ln}(L_y)$ as follows: $g = g(\infty) + \frac{a}{\ln(L_y) + b}$.}
 \label{fig:g-logLy}
\end{figure}

In particular, when the system temperature $T \ll T_c$, the convergence and accuracy of numerical calculations will significantly decrease due to the excessively large entanglement entropy, meaning that a large truncation dimension $D$ is needed. Therefore, we investigate the effects of bond dimension $D$ in Fig.~\ref{D56},  where two bond dimensions $D=200,280$ are compared in the five- and six-state Potts models. 

The Klein bottle ratio $g$ increases monotonically with $L_y$ increasing when $D=280$, however, $g$ calculated from $D=200$ exhibits oscillations and lie below the curve of $D=280$. It shows that $D=200$ is not enough for the characterization of $S_\text{KB}$, and, is consistent with the results in Ref.~\cite{ueda2026}, where the larger but not enough bond dimension $\chi$ renders larger deviation from the theoretical value with $\beta$ ($L_y$ in this paper) increasing. For larger $q$ or larger $L_y$, if the retained bond dimension is insufficient, $g$ tends to saturate to some stable value with some oscillations, as is shown in Fig.~\ref{fig:KBR_all}.

\subsection{What is $g$}

Unraveling the fundamental distinctions between the four- and five-state Potts models---and more broadly, characterizing continuous versus weakly first-order phase transitions through the lens of the Klein bottle ratio $g$---remains a compelling problem. Whether viewed as a universal CFT probe or a numerical diagnostic, $g$ consistently provides rich physical insights that extend beyond standard CFT paradigms.

\begin{figure}[t]
\centering
\begin{minipage}{\linewidth}
\centering
\includegraphics[width=\linewidth]{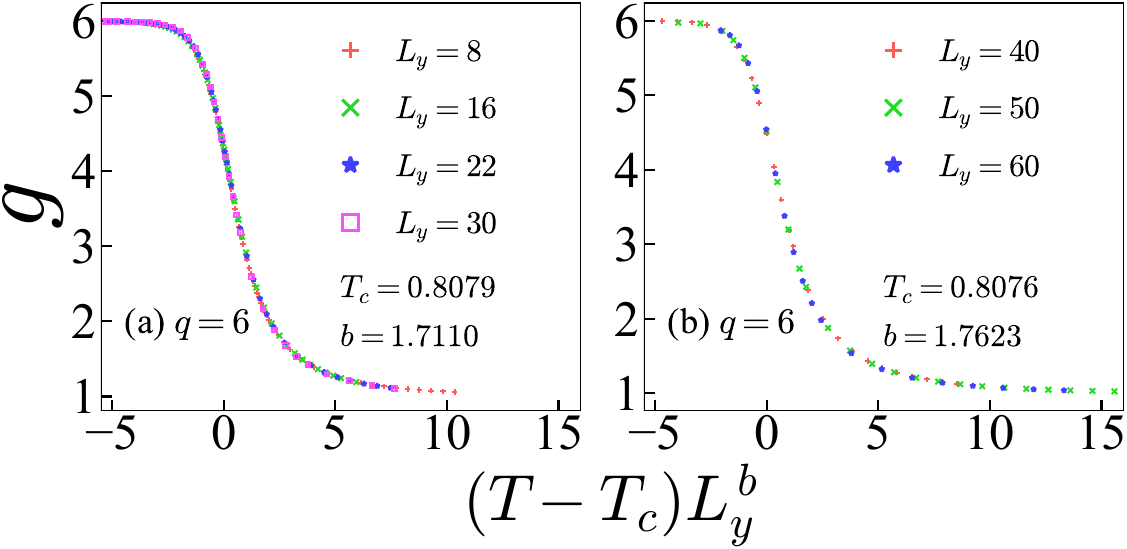}
\end{minipage}
\hfill
\caption{
Data collapse of the Klein bottle ratio $g$ for the six-state Potts model, comparing smaller-$L_y$ and larger-$L_y$ subsets at bond dimension $D=280$. (a) Fit for the smaller-$L_y$ subset: $T_c = 0.8079$, $b = 1.7110$. (b) Fit for the larger-$L_y$ subset: $T_c = 0.8076$, $b = 1.7623$.}

\label{fig:q56}
\end{figure}

In Figs.~\ref{fig:g-Ly} and \ref{fig:g-logLy}, the same dataset is fitted against different horizontal variables ($1/L_y$ and $\ln L_y$). For the three-state Potts model, the extrapolated ratio $g$ matches the exact theoretical prediction $\sqrt{3+6/\sqrt{5}} \approx 2.38396$. For the four-state Potts model, a traditional polynomial fit against $1/L_y$ [Fig.~\ref{fig:g-Ly}(b)] yields $g = 3.123$, consistent with previous work\cite{tang2017}. 

However, this value still deviates from the exact CFT prediction $g = \frac{9}{2\sqrt{2}} \approx 3.182$. In contrast, fitting $g$ against $\ln L_y$ [Fig.~\ref{fig:g-logLy}(b)] markedly improves the result to $g = 3.203$, providing direct and convincing numerical evidence for the logarithmic corrections induced by the marginally irrelevant operator\cite{difrancesco1997,tang2017,Morita2026}. Finally, for the five- and six-state Potts models, $g$ exhibits divergent scaling behavior, strongly confirming the first-order nature of the phase transition in the five-state model.

For the five- and six-state Potts models, our numerical calculations show that, even with a relatively large bond dimension $D=280$, data obtained across different system sizes $L_y$ still exhibit a surprisingly good data collapse (Figs.~\ref{b-drift} and \ref{fig:q56}). Furthermore, fitting $g$ accurately pinpoints the critical temperature, despite an exponent drift in $b$ similar to that observed for $q=5$. This behavior is intriguing, as the phase transition in the six-state Potts model is well established to be first-order, where physical observables are generally not expected to obey simple critical scaling relations. The observed data collapse therefore suggests that the Klein bottle ratio $g$ may possess unique or unconventional scaling properties. Nevertheless, larger system sizes must be carefully examined to draw definitive thermodynamic conclusions.

\subsection{Central charge}

\begin{figure}[t]
    \centering
    \includegraphics[width=0.99\linewidth]{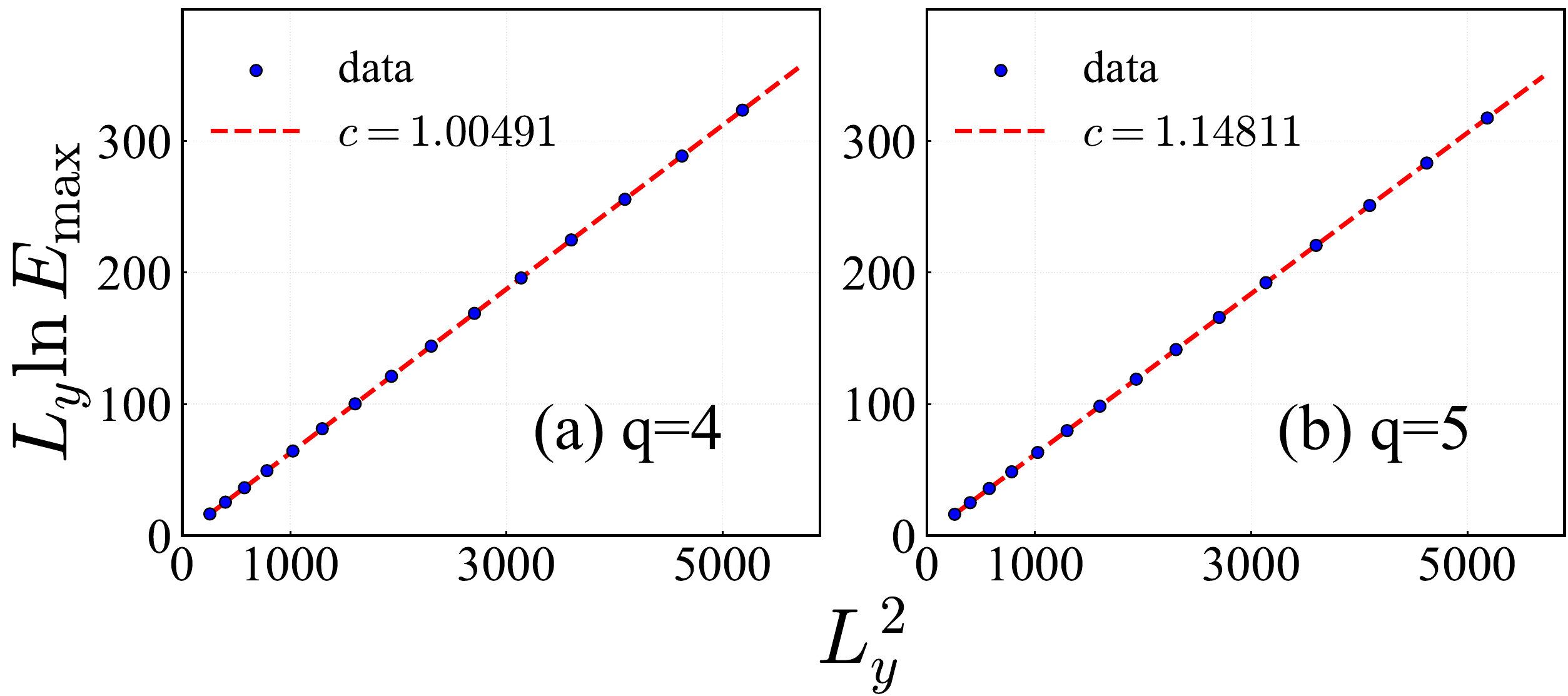} 
    \caption{The central charge $c$ extracted from the torus free-energy scaling for (a) $q=4$ ($c=1.00491$) and (b) $q=5$ ($c=1.14811$). Here, $D=280$.}
    \label{fig:central-charge}
\end{figure}

Combining with Eq.~(\ref{eq:ZT-ZK}), the central charge can be extracted by the intercept of the linear fitting of  $L_y \ln E_{max} ~ =-f_0 L_y^2 +\pi c/6$, as is shown in Fig.~\ref{fig:central-charge}. When $q=4$, the fitting result is $c = 1.00491$, consistent with $c=1$ from the compactified free boson CFT. For $q=5$, $c = 1.14811$. From the perspective of weakly first-order phase transition, it is beyond the description of conventional CFT. The calculated central charge is very close to the real part of theoretical prediction $c\approx 1.1375 \pm  0.0211i$ (with one more decimal by the analytical expression in Ref.\cite{tang2024}), characterized by the complex CFTs\cite{GRZ-SciPost}. Reference~\cite{ma2019} notes that weakly first-order phase transitions are proximate to complex fixed points, consequently manifesting a small imaginary component, which leads to the ``walking '' RG behavior and critical slowing-down\cite{GRZ-SciPost,GRZ-jhep}. 

\section{Conclusion}
\label{conclusion}

In summary, by using density matrix and tensor network algorithms we calculate the Klein bottle ratio $g$ to investigate the phase transition characteristic of the two-dimensional five-state ferromagnetic Potts model on square lattice. Through data collapse analysis of the Potts models for $q=4,5,6$, we find that the five-state Potts model exhibits significant pseudo-critical scaling due to its extremely large correlation length, while the critical exponent $b$ also shows the prominent size-dependence. It is consistent with the critical exponent drifting in weakly first-order phase transition. The underlying theory is that the complex fixed points lead to the ``walking'' behaviors in the real parameter space. 

For the four-state Potts model, we observe stable scaling behavior with an effective central charge of $c = 1.00491$, which is in excellent agreement with $c=1$ compact free boson CFT. The CFT predictions for the entanglement spectrum are further verified in Figs.~\ref{fig:sM} and \ref{fig:nLam}. Furthermore, by fitting $g$ as a function of $\ln L$, we quantitatively identify the logarithmic corrections induced by the marginally irrelevant operator\cite{salas1997,tang2017}. In contrast, the five-state Potts model exhibits an effective central charge $c = 1.14811$, closely matching the real part of the analytically continued value $c_{\text{Potts-}5} = 1.1375 \pm 0.0211 i$ predicted by complex CFTs\cite{GRZ-SciPost}.

We utilize the Klein bottle ratio, $g$, to analyze the phase transitions of two-dimensional five-state Potts model and find the scaling exponent $b$ drifting for the first time. 
We demonstrate that $g$ effectively locates the critical points for continuous ($q=3, 4$), weakly first-order ($q=5$), and even strongly first-order ($q=6$) transitions. For completeness, a detailed analysis of the $q=2$ and $3$ Potts models is presented in the Appendix, where CFT predictions are systematically benchmarked against numerical results. Although the scaling exponent $b$ drifts with system size for $q=5,6$ Potts models, finite-size scaling remains valid to pinpoint the critical point. In the low-temperature limit, $g=q$, which is just the ground-state degeneracy\cite{Affleck1991}. While $g$ was originally introduced within the context of CFT, its effectiveness in analyzing first-order transitions---which lies beyond standard CFT---suggests that $g$ encodes deeper physical information that warrants further investigation.

\subsection*{Acknowledgments}
We extend our gratitude to Hong-Hao Tu for his inspiring discussions and meticulous review of the manuscript. Wei Zhu is acknowledged for providing crucial interpretations of their work. We are grateful to Hai-Jun Liao for generously sharing the initial codebase. We also thank You-Jin Deng for insightful discussions during the project's early development. Finally, we are grateful to Zhi-Yuan Xie for his consistent and instrumental assistance with the tensor network algorithms. This work was supported by the National Natural Science Foundation of China (Grants No.~11874095 and No.~12047564).

\appendix
\counterwithin{figure}{section}
\renewcommand{\thefigure}{\thesection\arabic{figure}}

\section{the finite-size effect of the critical $q=2$ and $3$ Potts model}
\label{apsec:q23}

While the four-state Potts model exhibits prominent logarithmic corrections, the critical $q=2$ and $3$ Potts models correspond to unitary minimal models in two-dimensional CFT. Therefore, conducting a systematic analysis of finite-size effects across various $q$-state Potts models is highly instructive. From the renormalization group (RG) perspective, the finite-size corrections for $q=2$ and $3$ both exhibit power-law decay behaviors.

This behavior is closely connected to recent work\cite{2601.21502}. As shown in Fig.~\ref{fig:KBR_all}, the finite-size corrections for $q=2$ converge significantly faster than those for $q=3$. This rapid convergence stems from the fact that the finite-size scaling for $q=2$ exhibits an exponential decay governed by an exact analytical expression\cite{2601.21502}: $\sqrt{g} = \sqrt{2}\cos (\frac{\pi}{8}+\frac{1}{2}\delta)$ with $\delta = \arctan (e^{-\alpha L_y})$. For $L_y \gg 1$, the asymptotic expansion reads:
\begin{equation}
\sqrt{g} = \sqrt{\frac{2+\sqrt{2}}{2}} -\frac{1}{2}\sqrt{\frac{2-\sqrt{2}}{2}}e^{-\alpha L_y} + \cdots
\label{eq:g-asymp}
\end{equation}
where $\alpha = \frac{1}{2} \ln\left[\frac{\cosh{(2\beta)}+1}{\cosh(2\beta)-1}\right]$ and $\beta=1/T$. In the thermodynamic limit ($L_y \to \infty$), $\sqrt{g}=\sqrt{\frac{2+\sqrt{2}}{2}}$. Consequently, the quantity on the vertical axis, $\ln|\sqrt{g_{L_y}}-\sqrt{g_{\infty}}|$, serves as a linear measure of the deviation from the thermodynamic limit.

\begin{figure}[t]
\centering
\begin{minipage}{\linewidth}
\centering
\includegraphics[width=\linewidth]{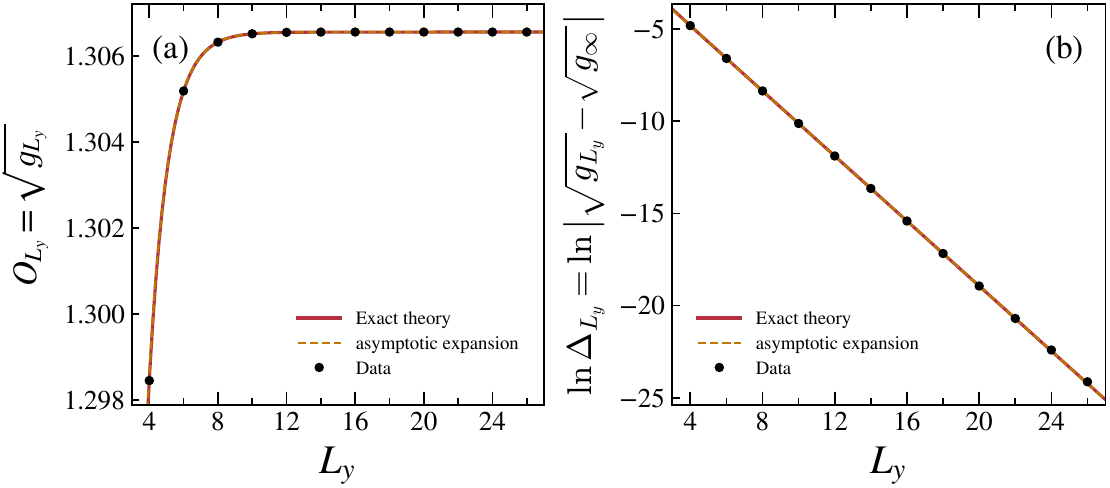}
\end{minipage}
\hfill
\caption{The two-state Potts (Ising) model:
(a) $\sqrt{g}$ as a function of system width $L_y$, comparing numerical data with the exact theoretical result and the asymptotic expansion [Eq.~\eqref{eq:g-asymp}].
(b) The leading exponential-decay correction, highlighted by plotting $\ln|\sqrt{g_{L_y}} - \sqrt{g_{\infty}}|$ versus $L_y$.}
\label{fig:q2exp}
\end{figure}

    As shown in Fig.~\ref{fig:q2exp}, the exact theory, its asymptotic expansion, and our numerical data exhibit remarkable agreement, collapsing seamlessly onto a single curve. Here, $g_{L_y}$ denotes the Klein bottle ratio $g$ for a system of width $L_y$. The exponential decay is clearly highlighted by the perfect linear behavior when plotting the logarithmic deviation of $\sqrt{g_{L_y}}$ from $\sqrt{g_{\infty}}$. 
    
\begin{figure}[t]
\centering
\begin{minipage}{\linewidth}
\centering
\includegraphics[width=\linewidth]{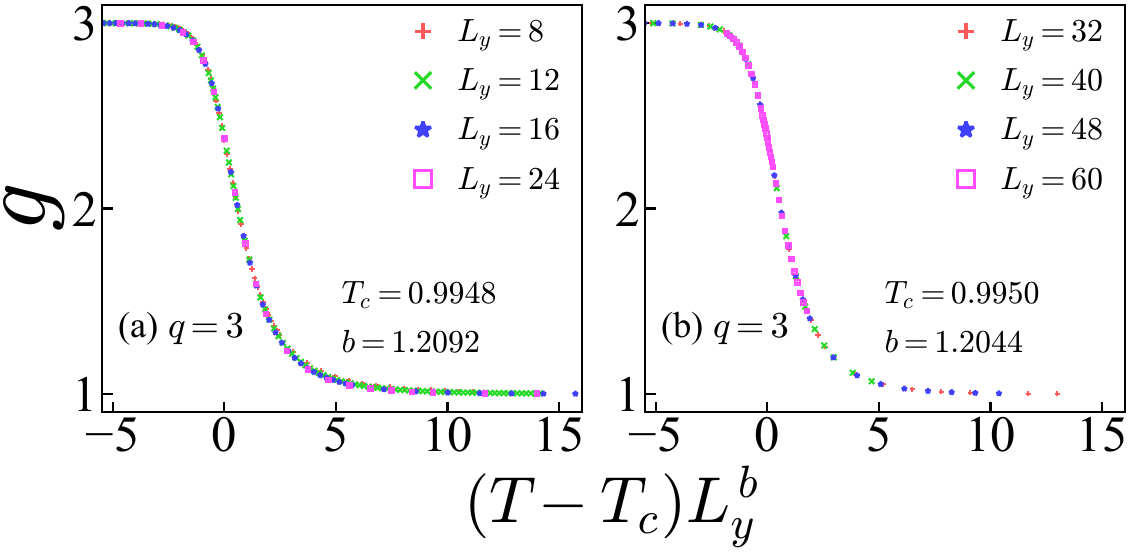}
\end{minipage}
\hfill
\caption{The three-state Potts model: Data from both the smaller-$L_y$ and larger-$L_y$ subsets are well fitted using critical temperatures $T_c = 0.9948$ and $0.9950$, respectively, in excellent agreement with the exact theoretical value $T_c^{\text{exact}} = 1/\ln(1+\sqrt{3}) \approx 0.994973$. (a) The fitted parameter value is $b = 1.2092$. (b) The fitted parameter value is $b = 1.2044$. 
}
\label{fig:q3}
\end{figure}

    In contrast, the finite-size corrections for the $q=3$ Potts model follow a power-law decay. As shown in Figs.~\ref{fig:g-Ly}(a) and \ref{fig:g-logLy}(a), the fitted coupling $g$ is very close to the CFT prediction $g=\sqrt{3+6/\sqrt5}\approx2.384$. The maximum system size reached in our calculations is sufficiently large, ensuring that the fit remains robust regardless of whether a power-law or logarithmic ansatz is applied. Independently fitting two distinct system-size ranges---the smaller and larger $L_y$ subsets---yields consistent critical temperatures $T_c = 0.9948$ and $0.9950$, both in remarkable agreement with the exact theoretical value $T_c^{\text{exact}} = 1/\ln(1+\sqrt{3}) \approx 0.994973$. As shown in Fig.~\ref{fig:q3}, the fitted exponent $b$ remains stable across different system-size ranges: $b=1.2092$ for smaller $L_y$ subset and $b=1.2044$ for larger $L_y$ subset, both slightly higher than the CFT prediction $b = 2 - 2h = 6/5 = 1.2$\cite{zhang2023}. Here, $h=2/5$ denotes the scaling dimension of the energy density\cite{difrancesco1997}.

In summary, it is remarkable that the finite-size corrections for $q=2,3$, and $4$ ferromagnetic Potts models systematically manifest as exponential, power-law, and logarithmic forms, respectively. There may be more to uncover, which we leave for further investigation.

\bibliographystyle{apsrev4-2} 
\bibliography{Potts}
\end{document}